\documentclass[twoside,a4paper,11pt]{sea10}
\usepackage{graphicx}
\usepackage{hyperref}
\usepackage{movie15}
\begin{document}
\pagenumbering{arabic}
\pagestyle{myheadings}
\thispagestyle{empty}
{\flushleft\includegraphics[width=\textwidth,bb=58 650 590 680]{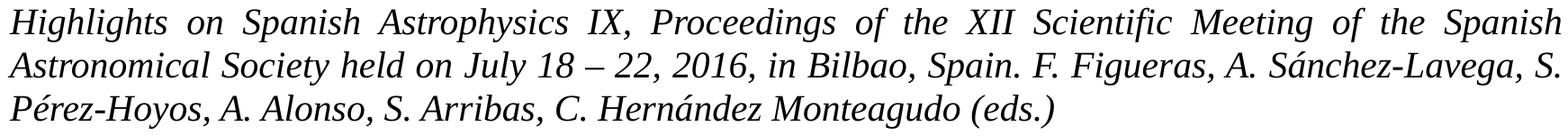}}
\vspace*{0.2cm}
\begin{flushleft}
{\bf {\LARGE
%
 Completing the puzzle:\\ AOLI full-commissioning fresh results and AIV innovations.
%
}\\
\vspace*{1cm}
%

Sergio Velasco$^{1,2}$,
Carlos Colodro-Conde$^{3}$,
Roberto L. L\'opez$^{1,2}$,
Alejandro Oscoz$^{1,2}$,
Juan J.F. Valdivia$^{4,5}$,
Rafael Rebolo$^{1,2,6}$,
Bruno Femen\'ia$^{7}$,
David L. King$^{8}$,
Lucas Labadie$^{9}$,
Craig Mackay$^{8}$,
Balaji Muthusubramanian$^{9}$,
Antonio P\'erez-Garrido$^{10}$,
Marta Puga$^{1,2}$,
Gustavo Rodr\'iguez-Coira$^{1,2}$,
Luis F. Rodr\'iguez-Ramos$^{1,2}$, 
and
Jos\'e M. Rodr\'iguez-Ramos$^{4,5,11}$ 
%
}\\
\vspace*{0.5cm}
%
$^{1}$
Instituto de Astrof\'isica de Canarias, c/V\'ia L\'actea s/n, La Laguna, E-38205, Spain\\
$^{2}$
Departamento de Astrof\'isica, Universidad de La Laguna, La Laguna, E-38200, Spain\\
$^{3}$
Departamento de Electr\'onica y Tecnolog\'ia de Computadoras, Universidad Polit\'ecnica de Cartagena, Campus Muralla del Mar, E-30202 Cartagena, Spain\\
$^{4}$
Departamento de Ingenieria Industrial, Universidad de La Laguna, La Laguna, Spain.\\
$^{5}$
Wooptix S.L., Torre Agust\'in Ar\'evalo, Avenida Trinidad, La Laguna, E-38205, Spain\\
$^{6}$
Consejo Superior de Investigaciones Cient\'ificas, Madrid, Spain\\
$^{7}$
W. M. Keck Observatory, 65-1120 Mamalahoa Hwy., Kamuela, HI 96743, Hawaii, USA\\
$^{8}$
Institute of Astronomy, University of Cambridge, Madingley Road, Cambridge CB3 0HA, United Kingdom\\
$^{9}$
I. Physikalsiches Institut, Universit\"{a}t zu K\"{o}ln, Z\"{u}lpicher Strasse 77, 50937 K\"{o}ln, Germany\\
$^{10}$
Departamento de F\'isica Aplicada, Universidad Polit\'ecnica de Cartagena, Cartagena, E-30202, Spain\\
$^{11}$
Centro de Investigaciones Biom\'edicas de Canarias, Campus Ciencias de La Salud s/n, E-38071 La Laguna, Spain

%
\end{flushleft}
%
\markboth{
Completing AOLI puzzle
}{ 
%
Velasco et al.
%
}
\thispagestyle{empty}
\vspace*{0.4cm}
\begin{minipage}[l]{0.09\textwidth}
\ 
\end{minipage}
\begin{minipage}[r]{0.9\textwidth}
\vspace{1cm}
\section*{Abstract}{\small
%
 The Adaptive Optics Lucky Imager (AOLI) is a new instrument designed to combine adaptive optics (AO) and lucky imaging (LI) techniques to deliver high spatial resolution in the visible, ~20 mas, from ground-based telescopes. Here we present details of the integration and verification phases explaining the defiance that we have faced and the innovative and versatile solution of modular integration for each of its subsystems that we have developed. Modularity seems a clue key for opto-mechanical integration success in the extremely-big telescopes era. We present here the very fresh preliminary results after its first fully-working observing run on the WHT. 
%
\normalsize}
\end{minipage}
%
%
%
\section{Introduction \label{intro}}

The Lucky Imaging (LI) technique, as suggested by \cite{1964JOSA} and named by \cite{1978JOSA}, was born as an alternative to AO to reach the diffraction limit in the optical bands. Images are taken at a very high frequency in order to select those intervals in which the atmosphere inside the collector tube through which the wavefront travels can be regarded as stable. If the best fraction of a bunch of images, those with smaller Strehl pattern, are stacked in a shift-and-add process, the equivalent to a high quality near-diffraction limit is obtained. The fraction of images that are selected for each target depends on the atmospheric conditions. The LI technique offers to ground-based telescopes an excellent and cheap method of reaching diffraction limited spatial resolution in the visible. 

One of the existing instruments to take advantage of the LI technique is FastCam, jointly developed by the Instituto de Astrof\'isica de Canarias (IAC) and the Universidad Polit\'ecnica de Cartagena (UPCT), described in \cite{2008SPIE}. FastCam, a common user instrument at the Carlos S\'anchez Telescope (CST, Teide Observatory, Canary Islands, Spain), routinely reaches the diffraction limit in the optical \textit{I} band both at the CST and the Nordic Optical Telescope (NOT, Roque de los Muchachos Observatory, Canary Islands, Spain), as on \cite{2011MNRAS}. In addition, FastCam has also obtained the image with the best resolution ever at the Canary Observatories (0.067” in \textit{I} band at the William Herschel Telescope, WHT, Roque de los Muchachos Observatory, Canary Islands, Spain), see \cite{2010SPIE}. 

However, this technique suffers two important limitations: it can be applied only at telescopes with sizes below 2.5m, achieving a resolution similar to that of the HST, and most of the images are discarded, meaning that only relatively bright targets can be observed. 

AOLI is a state-of-the-art instrument conceived to beat these limitations by combining the two most successful techniques to obtain extremely high resolution, LI and Adaptive Optics (AO). This instrument is hence planned as a double system that includes an adaptive optics closed loop corrective system before the science part of the instrument, this last using LI. The addition of low order AO with a new Two Pupil Plane Positions wavefront sensor (TP3-WFS) \cite{Colodro} to the system before the LI camera enhances the reachable resolution as it removes the highest scale turbulence maximizing the LI process at larger telescopes. 

Aiming at this challenging goal we have built AOLI, see \cite{VelascoSEA} and \cite{CraigSPIE}, putting together the expertise of several institutions -IAC, IoA, UPCT, UC and ULL-, each group specialized in a different subject, corresponding to a part of the puzzle.

\section{Instrument}

AOLI, still under commissioning, shall have the capability of using faint reference stars ({\it I\/} $\sim$17-18), reaching science targets as faint as {\it HST\/} but with a better resolution, meaning that the isoplanatic patch can be longer increased and that sky coverage available is much larger, even at high galactic latitudes. This will be thanks to its AO system, which consists of the TP3 wavefront sensor and a low order adaptive optics corrector using a deformable mirror together with photon counting EMCCD detectors.

  \begin{figure}[!h]
  \begin{center}
  \includegraphics[height=9cm]{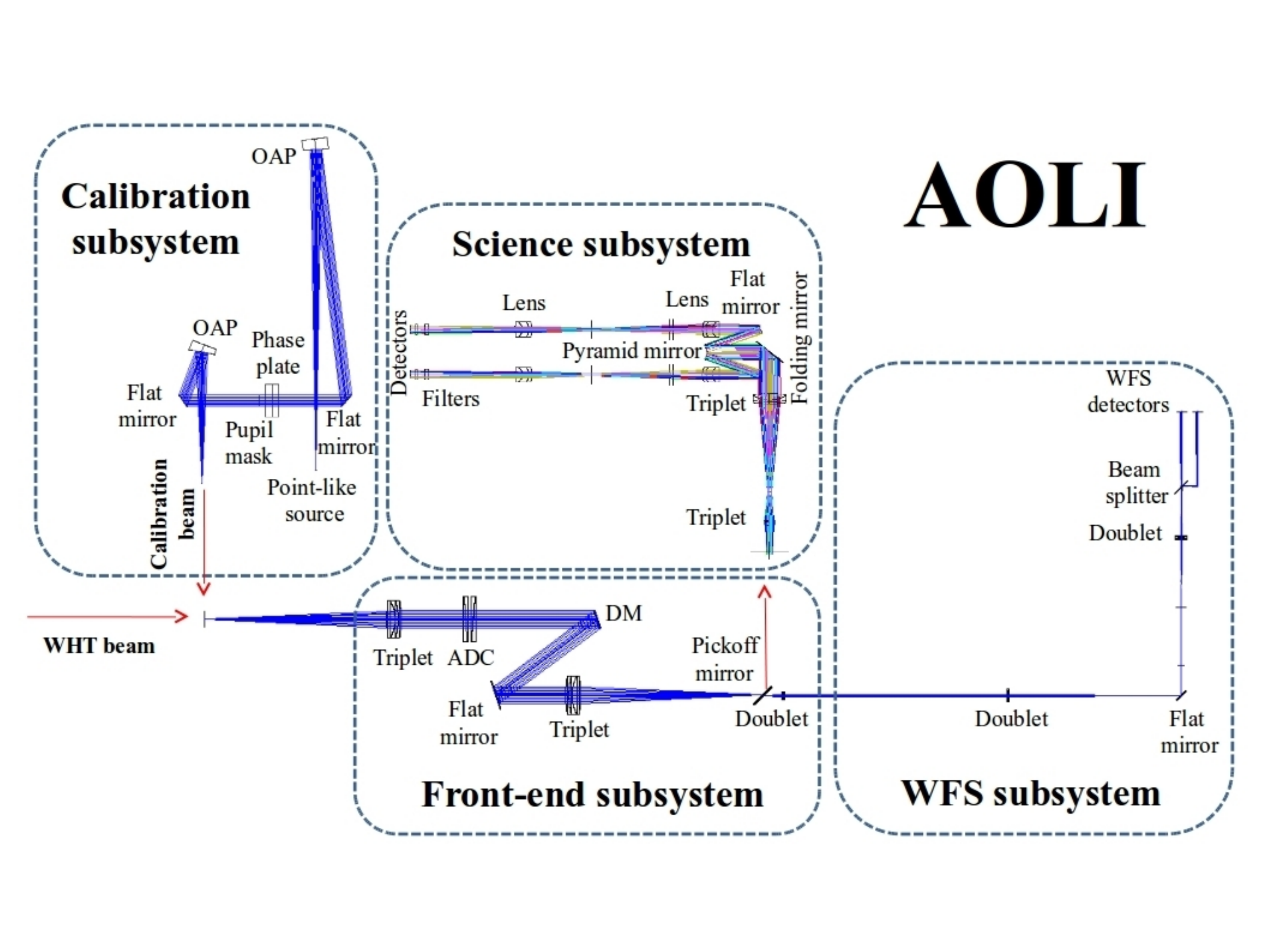}
   \end{center}
  \caption[example] { \label{fig:aoli} AOLI's optical layout. The AO subsystem before the science camera rises the amount of images that can be used for lucky imaging, allowing the observation of much fainter objects.}
  \end{figure} 

\subsection{Puzzle concept}

To face the defiance that AOLI represents we have implemented a new philosophy of instrumental prototyping by modularizing all its components: simulator/calibrator, deformable mirror (DM), science and WFS modules. This modular concept \cite{RobertoSPIE} offers huge flexibility for changes, such as the addition of future developments and improvements or the hosting telescope. AOLI has now been restructured not only to make the AIV phase reliable but also to be able to integrate this system regarding different parameters (f-number, scale, WFS-type,) or to adapt it to different telescopes. AOLI, initially designed for the 4.2m William Herschel Telescope (WHT, Observatorio del Roque de los Muchachos, La Palma island, Spain) can be adapted to other telescopes,  including the 10.4m GTC (ORM, La Palma island).

The four  modules that compound the system are: 
\begin{itemize}
\item Telescope and turbulence simulator and calibrator (SimCal): it delivers a calibrated point-like source resembling the telescope f-ratio and exit pupil and includes turbulence simulation of ground atmospheric layers.
\item Science subsystem based on four synchronized 1kx1k EMCCDs with three different plate scales and field of view (36x36 arcsec to 120x120 arcsec). 
\item Tomographic Pupil Image WFS (TPI-WFS), implemented for the first time on AOLI. It retrieves the pupil by measuring the intensity of defocused pupil images taken at two planes, allowing the use of much fainter reference stars than other WFS systems. The images are sent to two 512x512 pixels EMCCDs.
\item Front-end + AO subsystem: 241 actuators ALPAO deformable mirror (DM) and conditioning optics.
\end{itemize}

Besides the hardware, a unique sotware as of FastCam but scaled up, takes AOLI to a priority position. The possibility of processing the image on real time, even allowing the observer to check a live-view of it during the observation night not only becomes AOLI into a powerful scientific instrument but also reduces in a significant way the amount of data to be saved by computers. With an estimate of around two million images per night, this issue can not be undervalued.

\section{Commissioning}

On September 24th and 25th 2013, AOLI was installed at WHT's Nasmyth platform for a first commissioning to test its subsystems and overall efficiency without a fully developed AO subsystem. We were able to determine the plate scale (55.0$\pm$0.3 mas/pixel) and the PSF with a FWHM of 0.15 arcsec, probing them to be stable and to satisfy the specifications given. With just some seconds of on-sky integration we could probe the viability of the instrument offering some scientific results, see \cite{Velasco2016} and \cite{VelascoSEA16}.

\begin{figure} [ht]
  	\begin{center}
  		\includegraphics[height=4cm]{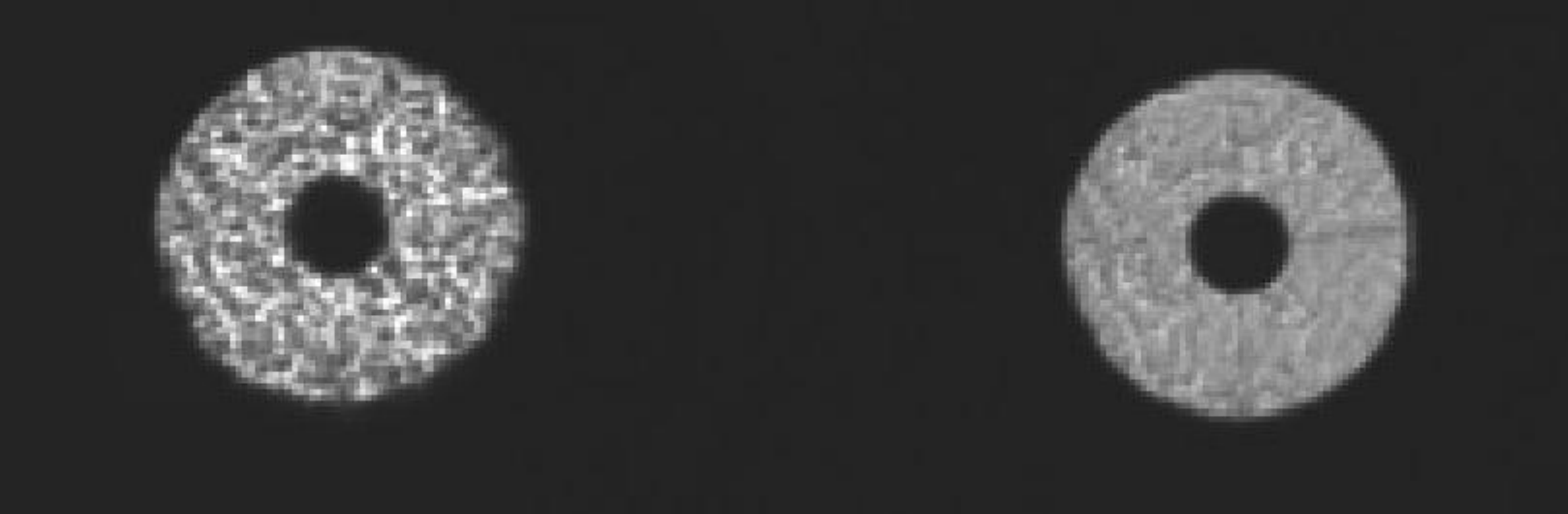}
  	\end{center}
  	\caption[example] { \label{fig:img6} TP3-WFS uses two defocused pupil images distant from the ideal pupil plane.}
  \end{figure}

\begin{figure} [ht]
  	\begin{center}
  		\includegraphics[height=6cm]{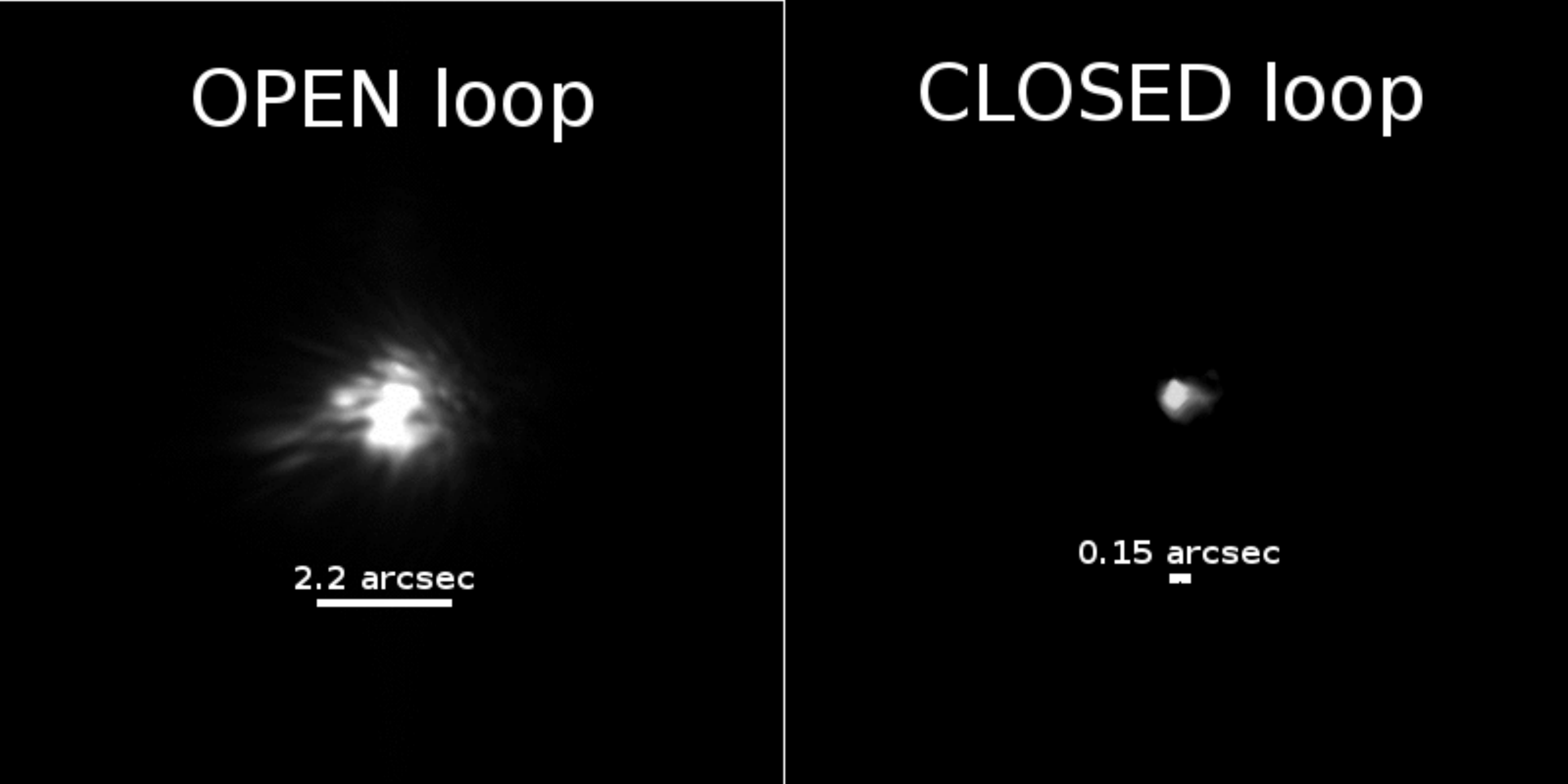}
   	\end{center}
  	\caption[example] { \label{fig:img7} The AO closed loop in the \textit{I} band.}
  \end{figure}

The first commissioning of the full instrument, including a fully working AO, took place at WHT on May 21st, 2016. The weather conditions that night were not optimal for high spatial resolution techniques due to a seeing around 2.2 arcsec, well above what is generally taken as a limit for AO systems, and full Moon. Despite of it, we closed the loop (see fig. \ref{fig:img7})  with different magnitudes stars making use of the TP3-WFS and an ALPAO deformable mirror with 244 actuators. In addition, the system was fed with a calibration made in real time from the image of the pupil (see fig. \ref{fig:img6}) from the reference star itself.

AOLI is, hence, the first instrument to succeed in closing the loop with stellar sources using the novel TP3-WFS. 

\small  
%

%

%
\end{document}